\documentclass[11pt,reqno]{article}
\usepackage{amsmath}
\usepackage{amsfonts}
\usepackage{amssymb}
\usepackage{hyperref}
\usepackage{latexsym}
\usepackage[dvips]{graphicx}
\usepackage{epsf}

\allowdisplaybreaks \numberwithin{equation}{section}

\newcommand{\be}{\begin{equation}}
\newcommand{\ee}{\end{equation}}
\newcommand{\bea}{\begin{eqnarray}}
\newcommand{\eea}{\end{eqnarray}}

\let\a=\alpha   \let\g=\gamma  
     \let\th=\theta  \let\k=\kappa \let\l=\lambda
\let\m=\mu    \let\n=\nu           
\let\s=\sigma     \let\ph=\phi 
    
 \let\D=\Delta   \let\L=\Lambda

\let\==\equiv

\newcommand{\p}{\partial}

\newcommand{\pr}[1]{ {}^{(#1)}\hspace{-0.05cm} }

\newcommand{\la}{\langle}
\newcommand{\ra}{\rangle}

\newcommand{\diff}{\chi} %symbol for discrete diffusion constant

\begin{document}
%------------------------------------------------------------------------------
\thispagestyle{empty}
\bigskip

\begin{center}
 {\LARGE\bfseries Spectral geometry as a probe of quantum spacetime}
\\[10mm]
Dario Benedetti, Joe Henson\\[3mm]
%$^1$
%$^1$
{\small\slshape
Perimeter Institute for Theoretical Physics \\
31 Caroline St.\ N, N2L 2Y5, Waterloo ON, Canada }\\
%{\upshape\ttfamily dbenedetti@perimeterinstitute.ca} } \\[3mm]

\end{center}
\vspace{5mm}

\hrule\bigskip

\centerline{\bfseries Abstract} \medskip
\noindent Employing standard results from spectral geometry, we provide strong
evidence that in the classical limit the ground state of
three-dimensional causal dynamical triangulations is de Sitter
spacetime. This result is obtained by measuring the expectation value
of the spectral dimension on the ensemble of geometries defined by
these models, and comparing its large scale behaviour to that of a
sphere (Euclidean de Sitter). From the same measurement we are also
able to confirm the phenomenon of dynamical dimensional reduction
observed in this and other approaches to quantum gravity -- the first
time this has been done for three-dimensional causal
dynamical triangulations. In this case, the value for the short-scale
limit of the spectral dimension that we find is approximately 2. We
comment on the relevance of these results for the comparison to
asymptotic safety and Ho\v{r}ava-Lifshitz gravity, among other
approaches to quantum gravity.
\bigskip
\hrule\bigskip

%------------------------------------------------------------------------------
\section{Introduction}
\label{sec:intro}
%------------------------------------------------------------------------------
%
The path integral approach to the non-perturbative quantisation of gravity has made unprecedented progress in recent years.
While a Euclidean path integral approach was suggested and studied many years ago \cite{Hawking:1980gf}, most of the non-perturbative results have remained at a formal level, because of the ill-defined nature of the functional integral.
Discretisation offers a way to better define such a path integral, as a continuum limit of a ``lattice'' theory, in analogy to standard lattice quantum field theory.  The causal dynamical triangulation (CDT) approach to discretisation \cite{Ambjorn:2009ts,Ambjorn:2005qt} has proven to be very useful in this regard.  In particular, it has allowed the calculation of expectation values of physically relevant observables as continuum limits of discrete observables.  These results provide some evidence that the CDT theory is approximating general relativity at large scales, the first step to confirming that a theory of quantum gravity has been discovered.

In particular, several results are consistent with the conjecture that the CDT theory is able to produce average geometrical configurations that approximate to the de Sitter solution of general relativity at large scales.  This is consistent with the claim that CDT theory has a semi-classical regime with the expected behaviour.  Not long ago, results on four-dimensional CDT \cite{Ambjorn:2004qm} showed that the effective large-scale dimension of the universe was approximately four, a non-trivial result in view of the possibility of pathological geometries dominating the path integral (as in the previous Euclidean DT approach \cite{Ambjorn:book}).  Measuring the expectation value of the volume of spatial slices as a function of cosmological time shows a good match to de Sitter geometry \cite{Ambjorn:2008wc}.  Fluctuations about de Sitter space in the CDT model can also be matched by  a reduced ``minisuperspace'' quantum cosmology model derived from general relativity.  Such results give the first evidence that, not only is the theory producing a solution to general relativity, but also the dynamics is consistent with what we know of quantised general relativity.

Still, it is difficult to measure a large and general class of geometrical observables with the computational methods that are being applied.  The results confirming de Sitter space are so far limited to information on the size of spatial slices, and dimension.  The difficulty is in finding genuine geometric observables (in keeping with the symmetries of general relativity) which are relevant at large scales (in keeping with the coarse-graining that we understand to be necessary to obtain physical results in quantum field theories). One approach to this problem was the proposal in \cite{Henson:2009fy} of a coarse-graining scheme for CDT, but the application of the method to simulations has not yet been achieved.  It would be desirable to identify a set of observables that could be more easily measured in Monte Carlo simulations of the CDT model, and rich enough to fully characterise the geometry.  This would be necessary in order to have a complete test that the model can produce an approximation to de Sitter at large scales (and, very similarly, that the geometry in the classical limit $L_{\text{Planck}} \rightarrow 0$ is exactly de Sitter).

One of the dimension estimators, the ``spectral dimension'', gives a hint of how this might be done.  In \cite{Ambjorn:2005db} it is explained how a scaling effective dimension estimator can be derived from the spectrum of the Laplace-Beltrami operator on a geometry.  The spectral dimension is a function of a ``diffusion time'' parameter corresponding to the relevant scale, and the function is in 1-to-1 correspondence with the spectrum of the Laplacian.  Its values can be found by examining diffusion processes on the geometry, which can easily be discretised on a CDT.  Interestingly, the results that can be gained in this way go beyond the dimension; some geometries (such as the Euclideanised version of de Sitter, the sphere) are entirely characterised by this spectral information.  Thus, by a closer examination of this information, readily available from Monte Carlo simulations, it may be possible to find strong evidence for the emergence of de Sitter space.  Indeed, this could be as close to a necessary and sufficient test as it is possible to get, within the practical limitations of computer simulations.

As well as the large-scale considerations, it is interesting to consider results from spectral geometry at short scales.  In 4D, current results \cite{Ambjorn:2005db} suggest a kind of ``dynamical dimensional reduction'', in which the spectral dimension approaches 2 at short scales.  This gives an example of the highly non-trivial nature of the UV regime of quantum gravity, gives some hints of how gravity might be a consistent theory in the UV limit, and also provides an observable to compare to results from other approaches to quantum gravity.

These ideas are explained in more detail below.  Applying them to the 3D CDT model, we find that large scale spectral dimension function derived from simulations matches the expected classical function very well at large scales.  We also find that the discrepancy is reduced as we approach a continuum (classical) limit.  As well as this, a new dimension estimator, based on the scaling of the spectral dimension function with the number of simplices $N$ in the CDT, gives a result consistent with three, with good accuracy.

These ideas are applied to the 3D model of CDTs here, for two reasons.  The first is that the simpler model allows simulations of larger linear size than the 4D model, and provides a testing ground for the ideas, which seem to be equally applicable to the 4D case.  Secondly, measurements of the spectral dimension have not previously been made in the 3D case.  This is of interest in itself, as measurements of the small scale spectral dimension for models of different dimensionality may allow a better comparison between the CDT models and various other candidate theories for quantum gravity, some of which have predictions for the small scale spectral dimension.  Here, we find the small scale dimension of 3D CDTs to be approximately two, as for the 4D model.

After briefly reviewing the essential concepts of spectral geometry in Sec.~\ref{s:heat_trace}, and the basics of CDT in Sec.~\ref{cdt-survey}, we describe in Sec.~\ref{s:implementation} how the two are combined. Finally in Sec.~\ref{s:analysis} we present our results and discuss their consequences. After the Conclusions, two appendices are added with some supporting material.
 Our results lead to a strong confirmation of the de Sitter character of the large scale geometry and some interesting comparisons of quantum gravity theories in three dimensions.

%------------------------------------------------------------------------------
\section{The heat trace and geometry}
\label{s:heat_trace}
%------------------------------------------------------------------------------
%
First, we review some relevant results from spectral geometry.  Consider a $d$-dimensional closed Riemannian manifold $M$ with a smooth, fixed metric $g_{\mu\nu}$.  The \textit{heat kernel} $K(\xi_0,\xi, \sigma)$ is a function on $M\times M\times \mathbb{R}_+$, which solves the heat equation,
\begin{equation}
\label{e:diffusion}
\frac{\partial}{\partial \sigma} K_g(\xi_0,\xi, \sigma) + \Delta_g K_g(\xi_0,\xi, \sigma) =0  \; ,
\end{equation}
with initial condition
\begin{equation}
\label{e:init_condition}
K_g(\xi_0,\xi, \sigma=0) = \frac{\delta^d(\xi - \xi_0)}{\sqrt{|g(\xi)|}} \; .
\end{equation}
Here,  $\sigma$ is the ``diffusion time'' (not to be confused with any physical time when considering the manifold as a spacetime), and $ \Delta_g = -g^{\m\n}\nabla_\m\nabla_\n$ is the scalar Laplacian on $M$ (acting on $\xi$), where $\nabla_\m$ is the covariant derivative compatible with  $g_{\mu\nu}$. In mathematical terms, the heat kernel $K_g(\xi_0,\xi, \sigma)$ is the Green function of the heat equation, while its physical interpretation is as the probability density of diffusion from $\xi_0$ to $\xi$ in diffusion time $\sigma$. The solution of \eqref{e:diffusion} is formally given by $K=\langle \xi| e^{-\s \D}|\xi_0 \rangle$, or in terms of eigenvalues $\l_j$ and eigenfunctions $\ph_j(\xi)$ of $\D_g$
\be \label{eigenf-heat}
K_g(\xi_0,\xi, \sigma)=\sum_j e^{-\l_j \s} \ph_j(\xi) \ph^*_j(\xi_0)\ ,
\ee
where it has to be understood that in the case that the spectrum is continuous the sum would be replaced by an integral.

 From this, we define a function that is a generally covariant observable, the \textit{heat trace},
\begin{equation}
P_g(\sigma) = \frac{1}{V_g} \int_M d\xi \, \sqrt{|g(\xi)|} \, K_g(\xi,\xi, \s)  \; ,
\end{equation}
where $V_g=\int_M d\xi \, \sqrt{|g(\xi)|}$ is the volume of the manifold, a normalization factor that is introduced for convenience.  The heat trace is the probability density for returning to the starting point of diffusion at time $\sigma$, integrated over $M$.  It is determined by (and determines) the spectrum via
\begin{equation}
\label{e:trace}
P_g(\sigma) = \frac{1}{V_g} \sum_j e^{-\lambda_j \sigma}  \; ,
\end{equation}
where the eigenvalues of $\Delta_g$ are repeated according to their degeneracy.  From this we see that the value of the heat trace at diffusion time $\sigma$ is not significantly affected by eigenvalues much larger than $1/\sigma$.  This is the relationship between diffusion time and the scale being ``probed''.  The heat trace is also related to invariants of the geometry through the well-known \textit{heat-trace expansion} \cite{Vassilevich:2003xt},
\begin{equation}
\label{e:trace-exp}
P_g(\sigma) = \frac{1}{ (4 \pi \sigma) ^{d/2} \, V_g} \sum_{n=0}^{\infty} a_n \sigma^n \; ,
\end{equation}
where $a_n$ is a series of curvature invariants, the first three of which are
\begin{gather}
a_0= \int_M d\xi \, \sqrt{|g(\xi)|} \, , \; \; a_1=\frac{1}{6}  \int_M d\xi \, \sqrt{|g(\xi)|} R(\xi) \, \\
a_2= \frac{1}{360} \int_M d\xi \, \sqrt{|g(\xi)|} \left\{ 5R^2 - 2 R_{\mu\,\nu}R^{\mu\,\nu} + 2 R_{\mu\,\nu \rho \tau}R^{\mu\,\nu \rho \tau}\right\} \, ,
\end{gather}
where $R$, $R_{\mu\,\nu}$, and $R_{\mu\,\nu \rho \tau}$ are the scalar curvature, the Ricci and the Riemann tensors respectively.  Thus, for these smooth geometries,  the heat trace determines dimension, volume, average scalar curvature and other curvature invariants.  In fact, in the case of the 3D and 4D spheres, the heat trace completely determines the geometry: it is known that if $M$ is a closed, connected Riemannian manifold of
dimension $2\leq n \leq 6$ with smooth metric $g_{\mu \nu}$, and if $(M,g)$ has the same spectrum as the $n$-sphere $S^n$ with the standard metric, then $(M,g)$ is in fact isometric to $S^n$ \cite{Tanno1973}.  This is not true for all geometries, however.  If two geometries have the same heat trace function, they are called \textit{isospectral}.

The function
\begin{equation}
\label{e:spectral_dim}
d_s(\sigma) = -2 \frac{d \ln(P_g(\sigma))}{d \ln \sigma}
\end{equation}
can be used to find the dimension of the manifold.  Considering for example  a flat space, the heat trace reduces to
\begin{equation}
P_g(\sigma) = (4 \pi \sigma) ^{-d/2} \, ,
\end{equation}
and so we see that $d_s(\sigma)=d$ in this case.  In the general case,
\begin{equation}
\label{e:specdim_general}
d_s(\sigma) = d- 2 \frac{\sum_{n=1}^{\infty} n\, a_n \sigma^n}{\sum_{n=0}^{\infty} a_n \sigma^n} \, ,
\end{equation}
and we see that the spectral dimension reduces to the topological dimension only in the limit $\s\to 0$, with a slope determined by the total curvature.
At very large diffusion time, one can see from \eqref{e:trace} that the spectral dimension has an exponential falloff determined by the lowest eigenvalues.

Up to here we have recalled well-known facts about spectral properties of classical manifolds. In the context of quantum gravity we expect that at very short scales spacetime will not look classical.  It will instead be replaced by some structure which, although unknown, has for some time generally been known as ``spacetime foam". In order to make some sense of this general expectation, one has to study in detail some specific model of quantum gravity and try to find at least an effective description of what happens to spacetime when we probe it at shorter and shorter scales. In this vein, we need to find geometrical ``probes" from which to obtain some indications of what is going on at the Planck scale. One possibility is to take, as such a probe, the heat kernel trace (adapted to the quantum gravity model in consideration) and study its deviations from classicality. This way we can gain some interesting insight on the short scale effects and also test whether the model has good classical properties at large scales. In particular we can use \eqref{e:spectral_dim} to define an effective notion of dimension for the quantum geometry.  In this case, we expect that the spectral dimension function of a good quantum gravity theory will approximate well to that of a classical geometry at large diffusion times, while at small diffusion times there will be significant deviations.  Below, we match the average spectral dimension function we derive from CDT simulations to that of a candidate classical geometry at large diffusion times, and then examine the small diffusion time behaviour to find properties of the spacetime foam.

Some of these ideas have been applied to $d=4$ CDT models by Ambj\o rn et al. in \cite{Ambjorn:2005db,Ambjorn:2005qt}, using methods that we compare to our own in section \ref{s:ajl}. The results they found showed for the first time a scale-dependent effective dimension which agrees with the classical (topological) one at large scales but reduces to 2 at short scales. This dimensional reduction is suggestive of a picture of quantum gravity in which the theory self-regularizes its behaviour, making itself safe from the UV-catastrophe which is usually associated with quantum gravity. Indeed $d=2$ is the critical dimension of gravity, and the theory presents no problems in that case. Such a picture seems compatible with that of the ``asymptotic safety scenario" \cite{Weinberg:1980gg,Niedermaier:2006wt,Percacci:2007sz}, for which the existence of a non-trivial fixed-point of the renormalization group equations implies that the theory near the fixed-point behaves like a two-dimensional theory (as deduced by the anomalous dimension). The spectral dimension has also been derived in the asymptotic safety scenario \cite{Lauscher:2005qz} and the results agree with CDT for the $d=4$ case.

Finally the spectral dimension has been studied also in the context of spaces with quantum group symmetry \cite{Benedetti:2008gu}, in loop quantum gravity and spin foams \cite{Modesto:2008jz,Modesto:2009kq,Caravelli:2009gk}, in Ho\v{r}ava-Lifschitz gravity \cite{Horava:2009if}, and in the strong-coupling limit of the Wheeler-DeWitt equation \cite{Carlip:2009kf}. We will comment more on the relation among these results, and their relation to our new findings, in section~\ref{sec:comparison}.

%------------------------------------------------------------------------------
\section{A brief survey of CDT}
\label{cdt-survey}
%------------------------------------------------------------------------------

The CDT models are a concrete proposal to define a path integral for gravity. A motivation for seeking such a definition is of course the hope that gravity might make sense if defined non-perturbatively. The traditional way to pursue a non-perturbative evaluation of a path integral in quantum field theory is to replace the continuum spacetime with a fixed lattice, a procedure which allows one to do actual computations (typically numerical ones, via simulations) with an object, the path integral, which otherwise has no meaning. Then one tries to recover the continuum theory, heavily borrowing concepts and procedures from the theory of critical phenomena.

The CDT approach exactly parallels the usual lattice field theory with one fundamental difference: the fixed lattice is replaced by an ensemble of random triangulations.
This is required by the fact that gravity is a theory of dynamical geometry, with no background spacetime fixed a priori.

More concretely, one defines an ensemble of ``triangulations" to work with, a triangulation being defined by a simplicial manifold,  i.e. a collection of $d$-dimensional flat simplices (the generalization of triangles and tetrahedra) glued along their $(d-1)$-dimensional faces and such that the neighbourhood of any vertex is homeomorphic to a $d$-dimensional ball.  A dynamical triangulation is one in which all the simplices are taken to be equilateral.  In the simulations we usually work with dynamical triangulations having a fixed number of $d$-simplices $N$, which we will denote $T_N$. The ensemble of such triangulations $\{T_N\}$ is obtained by gluing the $N$ simplices in all possible ways allowed by the simplicial manifold condition.
Furthermore, to avoid the sick behaviour that was found in the old models of dynamical triangulations, CDT models have one further restriction on the ensemble: only triangulations with a global time foliation, with respect to which no topology change occurs, are allowed. For more details on the geometrical meaning of this restriction and on its implementation see \cite{Ambjorn:2001cv}.

Once the ensemble is specified one can construct the partition function (Euclidean version of the path integral) as
\be \label{Z}
Z = \sum_N \sum_{T_N} \tfrac{1}{C(T_N)}\, e^{-S(T_N)} \, ,
\ee
where $S(T_N)$ is the bare action, and $C(T_N)$ is the order of the automorphism group of $T_N$, a symmetry factor naturally appearing when summing over unlabeled triangulations. Since we wish to recover general relativity in the classical limit, it is customary to use as a bare action the Einstein-Hilbert action adapted to a simplicial manifold, which is known as the Regge action. On a dynamical triangulation, the Regge action takes the very simple form
\be \label{action}
S(T_N) = \k_d N - \k_{d-2} N_{d-2} \, ,
\ee
where $\k_d$ and $\k_{d-2}$ are two coupling constants depending on the cosmological and Newton's constant appearing in the Regge action, and $N_{d-2}$ is the number of $(d-2)$-dimensional subsimplices (also called bones or hinges).

In principle one could use a different action, with more parameters, but at this stage this would only complicate the analysis of the results, and in a minimalist attitude such a generalization of the CDT models is usually postponed till the moment (if ever) at which the model itself will ask for such an extension. For example in $3+1$ dimensions a new parameter has been introduced in the action, without which no physically interesting region would exist in the phase diagram \cite{Ambjorn:2005qt}.
Furthermore, we need to remember that, as a consequence of topological relations, only $d/2$ (for $d$ even) or $(d+1)/2$ (for $d$ odd) among the values $\{N_0,N_1,...N_{d-1},N\}$ are independent. Hence, for $d=3$ and $d=4$ only two of such variables are independent, and as a consequence, if we want to keep the action linear in $N_j$, we only have two coupling constants. The counting changes if some anisotropy is introduced in the model, by assuming that the ratio $\a=l_t^2/l_s^2$ between the lengths of time-like and space-like edges\footnote{We are using here a Lorentzian language even though our signature is Euclidean; this is possible as in CDT it is always clear which edges in the triangulations are to be thought as time-like edges, and a Wick rotation is possible, at least before the continuum limit is taken.} is different from one. In such case, one finds that additional variables are needed in order to keep track of the orientation of the subsimplices, and new topological relations are found too. The counting for the anisotropic models was carried out in \cite{Ambjorn:2001cv}, and one has that for $d=4$ there are 10 variables and 7 constraints, leaving 3 independent variables, a fact that was used in  \cite{Ambjorn:2005qt} to introduce the new parameter. In $d=3$ the situation is instead unchanged with respect to the isotropic case, as there are 5 constraints for 7 variables, and hence again only 2 independent variables. For this reason it does not make sense to introduce in 3D the analogue of the new parameter used in 4D.

In this paper we concentrate on the case $d=3$, for which very few analytical results are known \cite{Ambjorn:2001br,Benedetti:2007pp} because of the difficulty in solving statistical models in dimensions higher than two. Hence we will resort to the method of Monte Carlo simulations.
In the simulations we will use the topological constraints to trade the variable $N_1$ for $N_0$, which is easier to keep track of, and replace \eqref{action} for $d=3$  by
\be \label{action-3d}
S(T_N) = \k_3 N - \k_0 N_0 \, .
\ee
Furthermore, as we mentioned, in the computer simulations we work at fixed volume, and hence we replace \eqref{Z} by
\be \label{Z_N}
Z_N =  \sum_{T_N} \tfrac{1}{C(T_N)}\,e^{\k_0 N_0} \, ,
\ee
where we have made use of the simple form of the action \eqref{action-3d}. Note that the partition function $Z$ is the discrete Laplace transform of $Z_N$ with respect to $N$.
The expectation value of an observable $A$ is calculated as
\be
\la A \ra_N = \frac{1}{Z_N} \sum_{T_N} \tfrac{1}{C(T_N)}\,  e^{\k_0 N_0} A(T_N)\, ,
\ee
which is related to the expectation value as a function of $\k_3$ via
\be
\la A \ra = \frac{1}{Z} \sum_N e^{-\k_3 N} Z_N \la A \ra_N \, .
\ee

Note that all the quantities appearing in \eqref{action} are dimensionless. Dimensions can be reintroduced in terms of the edge length $a$ of the simplices, the equivalent of the standard lattice unit, which is necessary when talking about the continuum limit. An essential part of the continuum limit procedure is to have $N\to\infty$ and $a\to 0$ in such a way that the physical volume
\be \label{volume}
V \sim a^d N
\ee
remains finite. This implies that when we want to give dimension to a dimensionless quantity by multiplying it by $a^n$, in practice we multiply it by $N^{-n/d}$.
In the continuum limit we expect that large-scale observables will become independent of the cutoff $a$, hence we expect to see finite size scaling when working with simulations at sufficiently large $N$, $i.e.$ we expect that an observable $f(g_1,...,g_m;N)$, depending on a set of $m$ couplings or variables $\{g_i\}$, will satisfy
\be \label{fss}
f(g_1 N^{-\tfrac{n_1}{d}},...,g_m N^{-\tfrac{n_m}{d}};N) = f(g_1 N'^{-\tfrac{n_1}{d}},...,g_m N'^{-\tfrac{n_m}{d}};N')
\ee
The natural expectation would be that \eqref{volume} and \eqref{fss} hold with $\{n_i\}$ given by the expected length-dimension of the $\{g_i\}$, but this is not guaranteed a priori and it is instead used as a check of the good classical properties of the model, as we do in the following.

%------------------------------------------------------------------------------
\section{Application of spectral geometry to CDTs}
\label{s:implementation}
%------------------------------------------------------------------------------

The expectation value of the spectral dimension function for CDTs was found from Monte Carlo simulations.  Finite size scaling analysis is applied to the results, as in \cite{Ambjorn:2000dja}.  Some previously existing code for the Monte Carlo simulations (used in \cite{Ambjorn:2000dja}) was adapted for this purpose.  These simulations fixed the spacetime topology to $S^2 \times S^1$, \textit{i.e.} spherical spatial sections and cyclical time.  The cyclical time is merely a convenience; it has been seen that the simulations produce an extended ``universe'' part consistent with a 3-sphere, and a ``stem'' part in which spatial slices remain near the minimum set by the lattice scale.  The ``stem'' part is thought to be a discreteness artifact caused by the topology used in the simulation.  Values of $N$ up to a maximum of 200k (meaning $2 \times 10^5$) were studied, although some errors for the larger values of $N$ are greater since less configurations could be generated to be averaged over, within practical time constraints.

All simulations were carried out with coupling constant $\kappa_0=5$, in the ``extended phase'' of the CDT phase diagram, where previous evidence points to the emergence of well-behaved geometry.  The total number of time-steps was set to $T=96$.  We approach the continuum limit with $\k_0$ fixed, $N \rightarrow \infty$ and keeping the dimensionful volume $V$ fixed.  In this case the Planck length is fixed in terms of lattice units, and as a result the dimensionful Planck length must go to 0 in this limit, and so this represents a classical limit.  In order to keep the Planck length finite a different limit would have to be taken with $\k_0$ a function of $N$ \cite{Ambjorn:2009ts}.

The heat trace of a CDT configuration can be found by appropriately discretising the diffusion process.  This process uses a probability distribution defined over the set of simplices of the CDT.  The analog of the continuum diffusion equation (\ref{e:diffusion}) is
\begin{equation}
K_T(x,x_0,\sigma+1) = (1-\diff) K_T(x,x_0,\sigma) + \frac{\diff}{4} \sum_{x' \in G(x)} K_T(x',x_0,\sigma)
\end{equation}
where $x$ and $x'$ labels of simplices, $G(x)$ is the set of simplices glued to $x$ (of which there are 4 in the 3D case), and $\diff$ plays the role of a diffusion constant, which can be between $1$ and $0$.  The distribution is initially peaked at $K_T(x,x_0,0)=\delta_{x, x_0}$.  Similarly to the continuum case, for a particular triangulation $T$ the average probability of return is
\begin{equation}
P_T(\sigma) = \frac{1}{N(T)} \sum_{x \in T} K_T(x,x, \sigma)  \; .
\end{equation}
and the analog of the spectral dimension of eqn.(\ref{e:spectral_dim}), $d_s(\sigma,T)$, is defined by replacing the differential with a finite difference in the obvious way.  The Monte Carlo simulations estimate the ensemble average of this quantity,
\begin{equation}
\label{e:discrete_ensemble_spec_dim}
D_{s}(\sigma,N) \equiv \la d_s(\s)\ra_N = \frac{1}{Z_N} \sum_{T_N} \tfrac{1}{C(T_N)}\, e^{\k_0 N_0}  d_s(\sigma,T_N)  \; .
\end{equation}
The time normalisation makes no difference to our results at this stage.  Similarly the value of $\diff$ is set purely from practical considerations, to minimise discreteness artifacts while allowing diffusion to proceed fast enough to probe large scales within the possible number of diffusion steps.  As explained in \cite{Ambjorn:2005qt}, at very early times the value of $P_T$ has an oscillatory behaviour due to discreteness effects.  This is somewhat mitigated by reducing the value of $\diff$, which was chosen to be 0.8 for the simulations.  It is also possible to introduce a different rate of diffusion through spacelike and timelike triangles on which the simplices are glued, which amounts to a scaling of timelike and spacelike distances, but this was not done here for reasons explained below.

In principle $D_{s}$ would be calculated from (\ref{e:discrete_ensemble_spec_dim}) by carrying out the diffusion process for \textit{all} tetrahedra in each Monte Carlo configuration.  However, a good estimate of this result can be found by randomly sampling the tetrahedra $x$ at which to calculate the probability of return $K_T(x,x, \sigma)$.  One tetrahedron was used as starting point for each configuration, selected uniformly at random from all tetrahedra not in the stem, and this was found to give reasonably small errors.

Before the process was simulated on the CDT, the CDT structure was converted from the data structure optimised for Monte Carlo moves to one optimised for the diffusion, resulting in a considerable speed-up.  This allowed the diffusion process to be followed for of the order of $20\, 000$ steps for all configurations, in the order of hours of computer time, more than enough for our purposes.  The diffusion process was carried out with the stem excluded. The stem here is defined as including all simplices in ``small'' slices ($<20$ simplices), and slices in ranges that lie between two small slices, but do not include the maximal slice (which would indicate that this range is the universe part).  The value of $N$ used in scaling calculations should, strictly speaking, be altered to take this into account, but this was not found to alter results in most cases \footnote{The exception is the value of the scaling dimension given in equation (\ref{e:scaling_dim}).  There, the value given was calculated using the total number of simplices less the average number of simplices in the stem, which was found to alter the result slightly.  Elsewhere the values of $N$ used are the total numbers of simplices.}.

One of our aims here is to compare these results from CDT simulations to the de Sitter geometry that we expect, from previous studies \cite{Ambjorn:2000dja}, to find at large scales.  The first step, then, is to determine the spectral dimension function for de Sitter, or the Euclideanised version, the sphere.  However, in practice, there is a complication here:  the lattice structure of CDTs breaks the symmetry between space and time.  Therefore we should not expect the ratio between the lattice lengths in the space and time direction that is used to define the bare action (set to 1 in this case) to be the correct one to use when comparing to a continuum geometry like de Sitter.  We are at liberty to perform some global rescaling of the time $t=a i$, where $i$ is the discrete time step, as was done in \cite{Ambjorn:2009ts} for the case of time-dependence of the volume of the spatial slices in 4D.  In that case, the time $t$ was taken as a coordinate time so that the scaling to the ``true'' cosmological proper time could be expressed as $\tau = \sqrt{g_{t t}} \, t$.  This becomes important when considering how to compare to de Sitter.

In the case that the observable under investigation is the spectral dimension, for which we don't know the explicit dependence on the proper time,
there are two ways to proceed.  The first, roughly, is to scale the CDT before comparing to the sphere.  This means carrying out the diffusion process given above with different diffusion rates for spacelike and timelike faces of the simplices.  The hope would be to extract the spectral dimension plot for a sphere by appropriately setting the ratio of these diffusion constants.  However, there is a limit on how fast diffusion in the time direction can be made using this technique.  The reason is that no simplicial path across more than two time slices in a CDT traverses only spacelike triangles, as each tetrahedron has at most only one spacelike face.  Because of this, no matter how fast diffusion across spacelike triangles is, diffusion across many time slices will not become arbitrarily fast. It was found in simulations that the scaling made possible in this way was not sufficient to reach the sphere, and so only results with equal diffusion rates on timelike and spacelike edges are given below\footnote{However, it is interesting to note that this would not be a problem if the diffusion was carried out on the vertices of the CDT lattice itself rather than on the tetrahedra; in that case there are paths traversing many time-slices consisting only of timelike edges.}.

The second way to proceed is to scale the sphere we are comparing to, rather than the CDT itself.  That is, we carry out the diffusion process using equal diffusion constants on spacelike and timelike triangles, but scale the cosmological proper time in the sphere that we wish to compare to before computing its spectral dimension function.  This means that we must compare to the spectral dimension function for the following metric, which we call the ``stretched sphere'':
\be \label{e:ssphere_metric}
ds^2_{S^3_s} = r^2 ( s^2 d\psi^2 + \sin^2\psi\ (d\theta^2 + \sin^2\theta\ d\phi^2))\, ,
\ee
where $\psi,\theta\in [ 0,\pi]$ and $\phi\in[0,2\pi]$, and $s$ is the deformation parameter (for $s=1$ we have the standard metric on a sphere of radius $r$).  The spectrum for this geometry is derived in App.~\ref{a:rugby}, and the spectral dimension can be computed from it, summing numerically the series \eqref{spectral-series}.

Our second goal will be to investigate the short-scale behaviour of the spectral dimension function, in order to confirm in 3D the phenomenon of dynamical dimensional reduction observed in 4D \cite{Ambjorn:2005db,Ambjorn:2005qt}. Going beyond the qualitative result, it is interesting to obtain a quantitative estimate of the asymptotic value (for $\s\to 0$) of the spectral dimension in order to compare it with results from other approaches to quantum gravity.

%------------------------------------------------------------------------------
\section{Analysis of the results}
\label{s:analysis}
%------------------------------------------------------------------------------

\subsection{Dimension from Scaling}

Before comparing to the sphere, the first task is to verify that the expectation value of the spectral dimension function is scaling with $N$ as might be expected for an extended 3D geometry.  Examining the scaling of the diffusion time of the spectral dimension functions with $N$ gives a way to estimate the large-scale dimension, similarly to the scaling dimension derived from volume-volume correlations discussed in \cite{Ambjorn:2000dja}.  Here, on large scales (\textit{i.e.} large diffusion times) we expect the functions to scale according to
\begin{equation}
D_{s}(\sigma N^{-2/d},N) = D_{s}(\sigma N'^{-2/d} ,N') \, ,
\end{equation}
because the diffusion time should scale with the square of the linear scale, from dimensional considerations (see eqn.(\ref{e:diffusion})).
We do not expect such simple scaling at small diffusion time, as this is the regime in which quantum effects become important, and these are probably governed by an additional parameter, most likely the Planck length. Since we keep $\k_0$ fixed we know that the Planck length will remain fixed in units of the cut-off, and so we cannot apply the finite-size scaling reasoning of the end of Sec.~\ref{cdt-survey} at these short scales.

\begin{figure}[ht]
\centering \resizebox{6.8in}{4.5in}{\includegraphics{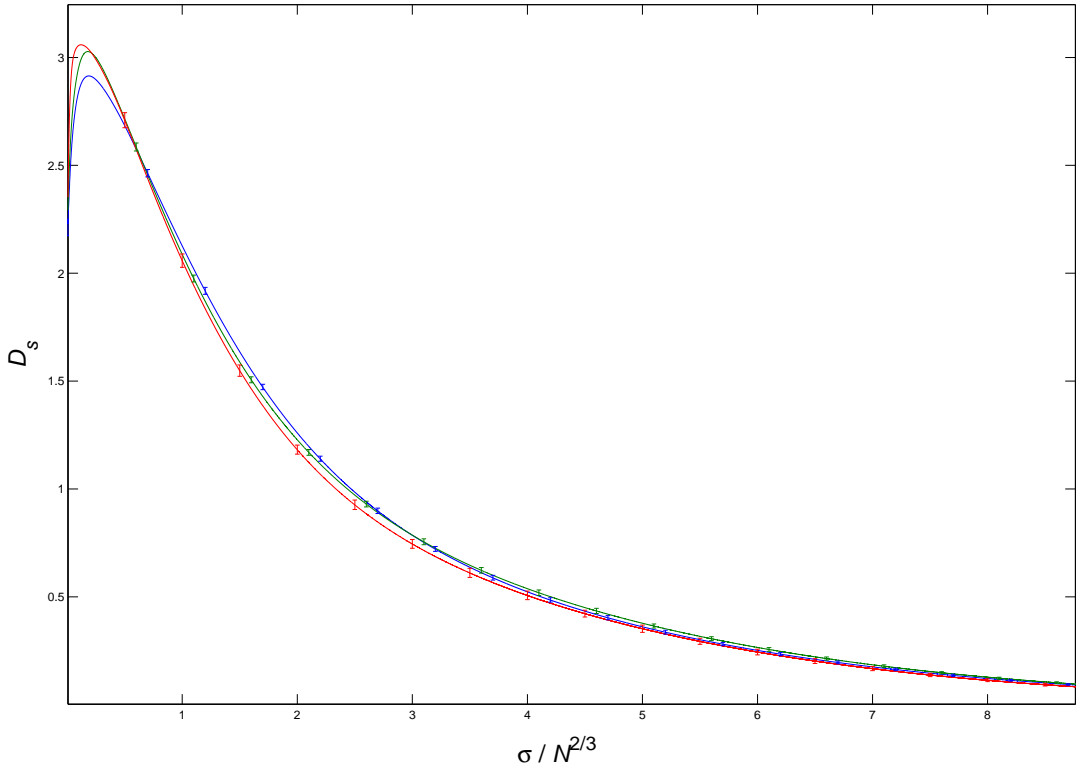}}
\caption{\small{The Spectral dimension data against scaled diffusion time $\tilde{\sigma}= \sigma N^{-2/3}$.  Data for $N=$50k, 100k and 200k is shown in blue, green and red respectively.  Error bars are suppressed on most data points for clarity (this is done throughout the paper). The curves should converge to a limiting curve as $N \rightarrow \infty$.  At these values of $N$ it there is good evidence of convergence to a continuum limit:  the curves for $N=$ 100k and $N=$ 200k agree within error for  $\tilde{\sigma} > 0.5$, as do the  $N=$ 50k and $N=$ 100k curves.
}\label{f:scaling}}
\end{figure}

 Figure \ref{f:scaling} shows that a good match between data sets with different value of $N$ is possible with $d=3$, and illustrates that the simulations are showing the expected convergence to a continuum limit at large diffusion times.  The best estimate of $d$ was found by comparing the $N=50$k and $N=100$k data sets using best overlap methods\footnote{Comparisons to $N=200$k data were also consistent with $d=3$ but the errors were larger due to larger Monte Carlo errors.}.  The range of times to be compared was fixed for the $N=100$k data to $2 < \sigma N^{-2/3}< 7.4$ to exclude short scale quantum effects.  The data was scaled appropriately for various values of $d$ and compared using linear interpolation.  The best overlap was found at
\begin{equation}
\label{e:scaling_dim}
d=2.99 \pm 0.12 ,
\end{equation}
where the errors are the Monte Carlo random errors (finite difference errors and so on being negligible compared to these).  This method gives an estimate of the dimension that improves on the accuracy of previous methods, and gives yet more evidence for the 3 dimensional nature of the geometry at large scales.  In view of this result, we define $\tilde{\sigma}= \sigma N^{-2/3}$ as an appropriately scaled diffusion time, so that we can compare simulations at all values of $N$ to a stretched sphere of fixed volume.

%%%%%%%%%%%%
\subsection{Comparison of CDT results to the stretched sphere}

\begin{figure}[ht]
\centering \resizebox{6.8in}{4.5in}{\includegraphics{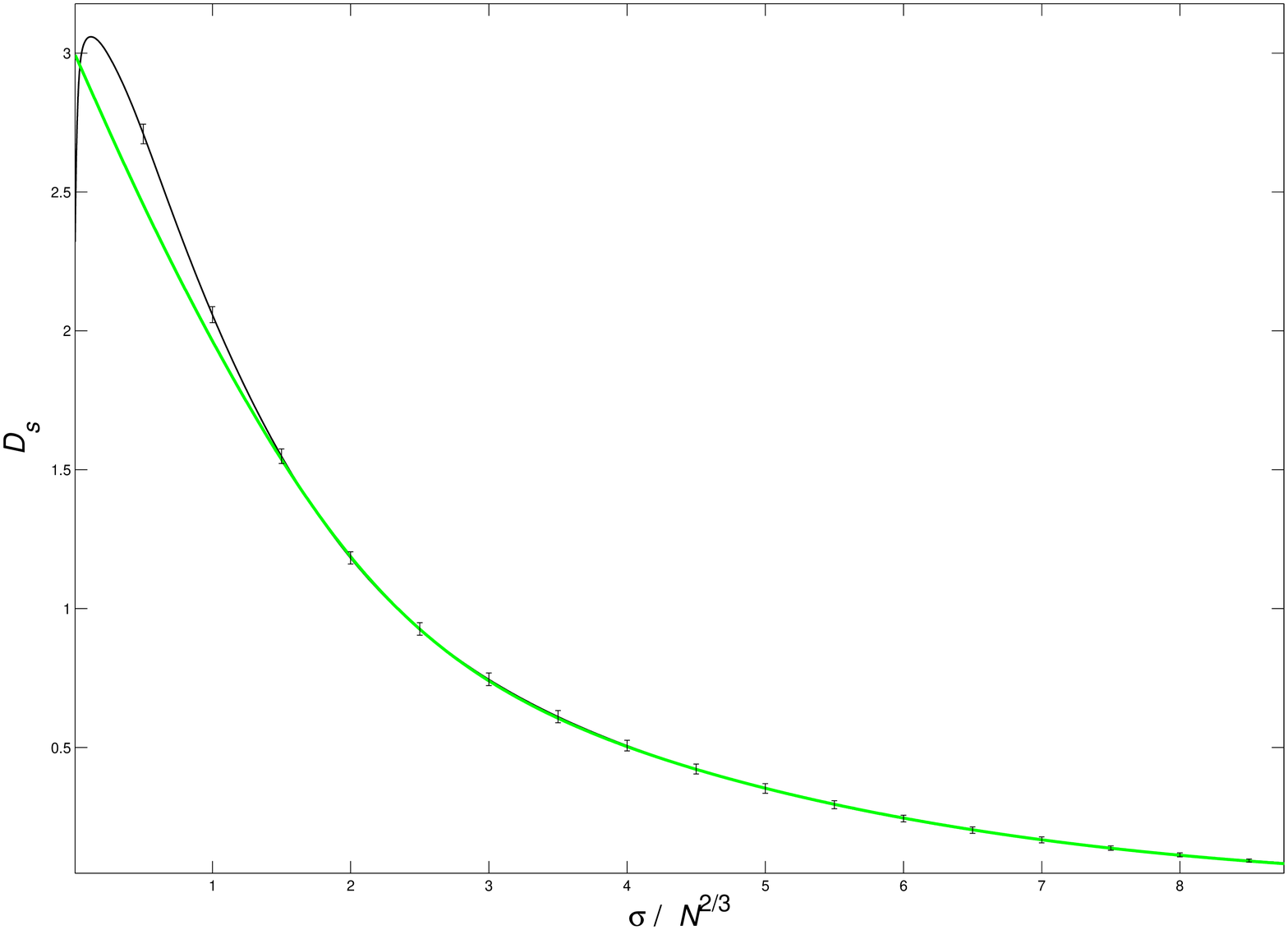}}
\caption{\small{The Spectral dimension data from CDT simulations with $N=200$k against scaled diffusion time $\tilde{\sigma}= \sigma N^{-2/3}$, plotted in black, fitted to the spectral dimension plot for the scaled sphere with $r=1.20$, $s=1.96$, superimposed in green.  The two-parameter fit agrees with the data for $\tilde{\sigma} \gtrsim 1.5$.  at very low times the ``quantum correction'' to the posited classical geometry is negative, but it becomes positive in an intermediate range.
}\label{f:spherematch}}
\end{figure}

\begin{figure}[ht]
\centering \resizebox{4.5in}{3.0in}{\includegraphics{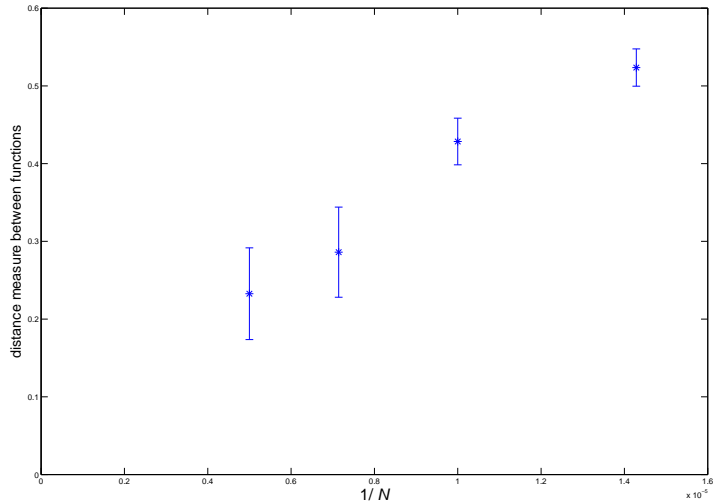}}
\caption{\small{The overlap between the CDT data for $D_s(\tilde{\sigma},N)$ and that for the scaled sphere with $r=1.20$, $a=1.96$, as a function of $1/N$, with errors.  Data for $N=70$k , 100k, 140k, and 200k is used here.  The difference between the two functions is measured as the integrated absolute difference between them in the range $0<\tilde{\sigma}<7.4$. As $1/N \rightarrow 0$ the difference comes closer to 0.  The results are consistent with the convergence of the CDT data with the scaled sphere spectral dimension function as $N \rightarrow \infty$.  (At smaller values of $N$ than 70k, there is competition between the positive and negative quantum corrections that obscures the convergence.)
}\label{f:fits_against_N}}
\end{figure}

Our main goal here is to determine whether the spectral dimension function for a stretched sphere can match that of eqn. (\ref{e:discrete_ensemble_spec_dim}) at large diffusion times.  Figure \ref{f:spherematch} shows that a scaled sphere with $r=1.20$, $s=1.96$ matches the $N=200$k Monte Carlo data for $\tilde{\sigma} \gtrsim 1.5$.  The good quality of the 2-parameter fit is evident from the plot.  Note that the quality of the fit is not generic in any sense: a simple exponential gives a bad fit over this range, and altering the value of $s$ by more than $0.01$ also makes a noticeable difference to the quality of the fit.   Evidence was also found that the spectral dimension function for CDTs converges to that for the scaled sphere as $N \rightarrow \infty$, as shown in figure \ref{f:fits_against_N}.

This remarkable fit not only shows consistency with the emergence of a 3-dimensional extended geometry on large scales, but also for the consistency between all the spectral properties of the scaled sphere and those of the results of the CDT simulations at large scales. In other words, these results provide strong evidence that the average CDT geometry for the largest simulations is isospectral to some geometry that approximates to the stretched sphere at large scales.  It also indicates that, in the classical limit, the geometry is isospectral to the scaled sphere.  We interpret this as the strongest evidence so far that the 3D CDT simulations are approximating 3D de Sitter space at large scales.

It is interesting to note the form of the quantum correction to the classical geometry, as given in figure \ref{f:spherematch}.  At very low times the spectral dimension for the CDT data is lower than the classical value, as found in the 4D case (this is investigated more fully in the next section).  However, there is an intermediate regime in which the quantum correction is positive, and persists for a longer diffusion time than might be expected from looking at the behaviour at very low diffusion times.

\subsection{Results at shorter scales}
\label{s:ajl}

These results have implications for previous methods used to measure the large scale dimension of CDTs.  In \cite{Ambjorn:2005db}, in the 4D case, Ambj\o rn, Jurkiewicz and Loll (AJL hereafter) take a short range of early diffusion times, and fit the CDT spectral dimension data to the function
\begin{equation}
D_s(\sigma) = a - \frac{b}{\sigma +c}\, .
\end{equation}
This fit provides estimates of the dimension as $\sigma \rightarrow 0$ and as $\sigma \rightarrow \infty$.  In particular, the estimation of the large scale dimension rests on two assumptions.  The first is that the finite size effects have a negligible effect on the fit over the range fitted.  The second is that the fit is good into the range at which the quantum corrections to the spectral dimension become negligible.  In other words, the classical spectral dimension function being limited to as $\sigma \rightarrow \infty$ is approximated as a constant function, and it is assumed that the quantum corrections affect only the non-constant term.  As $N$ is increased, and the range of the fit is kept as a constant range in $\sigma$, the (negative) finite size errors become smaller.  However the errors from the quantum deviations from the classical geometry will not go to zero unless these quantum deviations are well-fitted by $b/(\sigma +c)$.

\begin{figure}[ht]
\centering \resizebox{6.8in}{4.5in}{\includegraphics{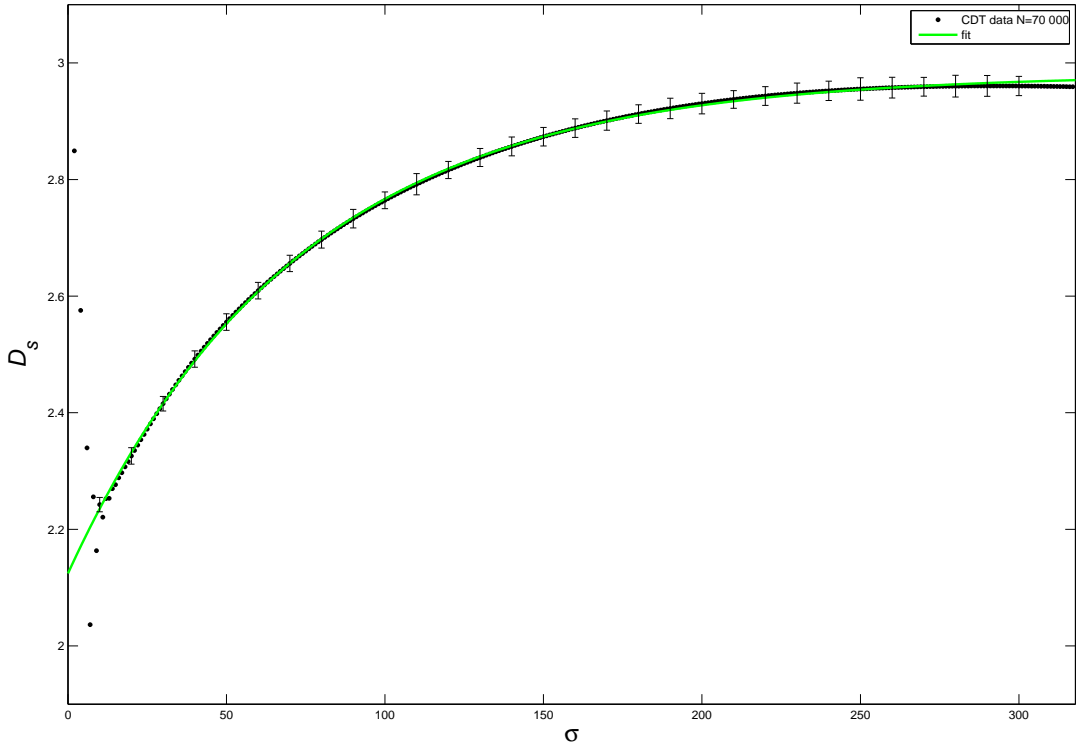}}
\caption{\small{The Spectral dimension data with number of simplices $N=70k$, at small diffusion times, is plotted with error bars in black.  The exponential fit is superimposed in a lighter colour.
}\label{f:70ksmall}}
\end{figure}

The results of the previous section show this latter assumption to be only approximately correct, in the 3D case at least.  To illustrate this further, we repeat an analysis similar to the AJL analysis.  At $N=70k$, we find that in the 3D case the data is similarly fitted with 3 parameters, with the large scale limit being a constant.  In this case, the following fit is better than the rational fit given above:
\begin{equation}
D_s(\sigma) = a + b e^{-cx}.
\end{equation}
The fit is taken between $\sigma=20$ and  $\sigma=300$ to avoid both short scale discreteness effects and finite size effects, as in \cite{Ambjorn:2005db}.  As can be seen from Fig. \ref{f:70ksmall}, the fit is good (although it matches less well at large and small times), and the estimates derived (with errors calculated combining Monte Carlo random errors with fitting errors and finite difference errors) are the following:
\begin{align}
D_s(\infty,N=70k) = 2.98  \pm 0.02 \; ; \\
D_s(0,N=70k) =  2.12  \pm 0.04 \; .
\end{align}

The large scale dimension is consistent with 3 here, even with the assumptions given above, and one expects the finite size error to be small and negative.  This is similar to the situation found by AJL in the 4D case, where the large scale dimension is consistent with 4.  However, as one goes to yet higher $N$, the technique begins to slightly overestimate the dimension in the 3D case.

\begin{figure}[ht]
\centering \resizebox{6.8in}{4.5in}{\includegraphics{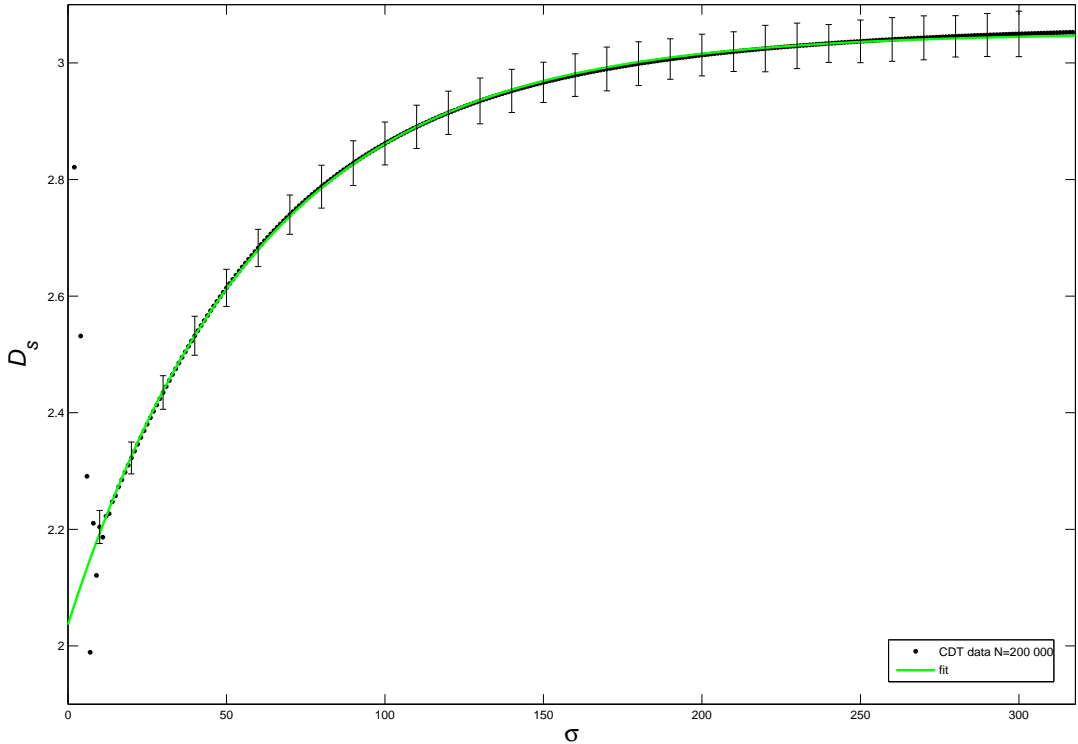}}
\caption{\small{The Spectral dimension data with number of simplices $N=200k$, at small diffusion times, is plotted with error bars in black.  The exponential fit is superimposed in a lighter colour.
}\label{f:200ksmall}}
\end{figure}

At $N = 200 k$ we have a similarly good fit with the same function (see Fig. \ref{f:200ksmall}).  The estimates in this case are
\begin{align}
D_s(\infty,N=200k) &= 3.05 \pm 0.04 \; ; \\
D_s(0,N=200k) &=  2.04  \pm 0.10 \;.
\end{align}

Here, the value 3 for the large scale dimension begins to looks doubtful.  This is explained by looking at the spectral dimension plot at larger scales, as in figure \ref{f:spherematch}.  There we see that the quantum correction to the spectral dimension function for the conjectured classical geometry actually becomes \textit{positive}, and stays significant for some time after the maximum value of $D_s$ is reached.  But the above fits are taken over a short range before the maximum is reached.  This means that the assumption that the fit will have a large scale limit free of quantum effects is not exact, giving a positive bias in $D_s(\infty)$.  However, the correction to the estimates given above because of this effect cannot be larger than the maximum quantum correction to the classical spectral dimension, which is 0.28 .  Thus, we see that the extra error that we should include is not so large that it totally destroys usefulness of the technique; even with the approximate assumption this method does show that the large-scale dimension converging to a value consistent with 3, with approximately 10\% error.  The dimension estimations of the previous section also give accurate new confirmations that the dimension converges to 3, with less assumptions.  The situation in 4D may well be similar.

It is interesting to note the possibility of measuring an effective average scalar curvature in a similar way. From eqn. (\ref{e:specdim_general}) we see that, classically, the gradient of the spectral dimension function as $\sigma \rightarrow 0$ is related to this observable.  We could define an effective scaling version of this observable, proportional to the gradient $d D_s(\sigma) / d \sigma$, and proceed in roughly the same way as in the AJL method of measuring the large scale dimension.  For this to give a meaningful value, the range of values of $\sigma$ at which only the first two terms of the heat trace (\ref{e:trace-exp}) are contributing to the spectral dimension must be appreciably longer than the range over which quantum effects are important.  From figure \ref{f:spherematch} we can see than our simulations do not allow this at present.  This would apply also to other ways of measuring curvture \cite{deBakker:1994zf,Henson:2009fy}.  This provides an interesting example of a situation in which concepts from general relativity and quantum field theory come together: although the average scalar curvature is defined as an integral of a function defined at each point, we know that this ``maximally fine-grained'' notion will no longer be the physically appropriate one in the quantum theory, and that we should only expect the concept of average scalar curvature to be useful in some coarse-grained sense, at some intermediate scales.  In any case, the fit of the spectral dimension function to the scaled sphere shown in figure \ref{f:spherematch} gives good evidence for consistency with de Sitter beyond just the dimension or average scalar curvature.

The two assumptions necessary to estimate the large scale dimension are not necessary to estimate the small scale dimension. The $N=200k$ simulations are the largest available, and give a value consistent with 2 here, but inconsistent with 1.5. We will comment extensively on this in the following subsection.

%------------------------------------------------------------------------------
\subsection{Discussion of the $\s\to 0$ limit}
\label{sec:comparison}
%------------------------------------------------------------------------------

It is interesting to compare the short-scale behaviour found here for the spectral dimension, to that derived from other approaches to quantum gravity.
In most of the cases we are unable to give the analytic dependence on the diffusion time, and hence we can only discuss the qualitative behaviour, and speculate about the limiting value $D_s(\s\to 0)$.
 While it would certainly be a bold extrapolation to claim that two approaches are describing the same physics just because they give the same number as  $\s\to 0$, it is intriguing to observe a certain universality of the results like it seems to happen in $d=4$ \cite{Carlip:2009kf}.
In this respect it is useful to carry out this type of study in general dimensions, as for example $d=4$ might be a special case in which several approaches agree on $D_s(\s\to 0)$ despite having some fundamental difference. In this sense we think that a discrepancy in the results might have a stronger meaning, as a disagreement would probably signal some fundamental difference, and hence the dimension-dependence of the results might teach us something important. Our interest in three-dimensional CDT originates mainly from these concerns.

We notice first of all that our result agrees with that found for $d=4$ in \cite{Ambjorn:2005db}, $i.e.$ we find $D_s(\s\to 0)\sim 2$ both in three and four dimensions. We should have a word of caution here, in light of the recent observation  \cite{Ambjorn:2009ts} that, at least in the 4D case, the Planck length is of order the lattice spacing $a$ at the currently possible parameter settings in Monte Carlo simulations, and remains fixed when the ``bare Newton coupling'' is fixed.  This is something that we have not tested in our 3D simulations, but which could well hold true in this case. This observation implies that we are at the moment unable to probe physics well beyond the Planck scale in a reliable way ($i.e.$ without discretization artifacts), hence the ``$\s\to 0$" limit should be seen as a simple extrapolation of the results at $\s>L_P^2$. In order to test whether such extrapolation is correct we would need to make the lattice length much smaller in comparison with the Planck scale, as described in \cite{Ambjorn:2009ts} for 4D. This will not be attempted here, however.
For the sake of discussion we will assume that we are seeing enough of the genuine UV behaviour at the values chosen for the parameters, and proceed to compare our CDT result to that coming from other approaches to 3D gravity. Further study will confirm whether this assumption is justified.

In \cite{Lauscher:2005qz} the general $\s\to 0$ limit of the spectral dimension for asymptotically safe gravity in $d$ dimensions was derived, resulting in $D_s(\s\to 0)=d/2$.
Interestingly, for $d=2$ this coincides with the AJL results, but for $d=3$ it disagrees with the results found here. It should be stressed that in CDT, in both cases, the dimension is evaluated numerically, hence we can't be sure it will be exactly an integer. On the other hand the small errors give us confidence that a value of  $D_s(\s\to 0)=3/2$ in $d=3$ can be excluded.

In $(n+1)$-dimensional Ho\v{r}ava-Lifshitz gravity with characteristic exponent $z$ one finds \cite{Horava:2009if} that $D_s(\s\to 0)=1+n/z$. Hence $D_s(\s\to 0)=2$ in the $(2+1)$-dimensional case\footnote{Note that $n=2$ here corresponds to what we have called $d=3$ in the rest of the paper.} with $z=2$, which is the $(2+1)$-dimensional analogue of the $(3+1)$-dimensional case with $z=3$ proposed in \cite{Horava:2009uw}, $n=z$ being the critical dimension of models characterized by $z$.
This coincidence, together with the observation made in App.~\ref{app:Lifshitz}, suggests that the link between CDT and Ho\v{r}ava-Lifshitz gravity deserves to be explored further.

In the context of spin foam models of three-dimensional quantum gravity, the results of \cite{Caravelli:2009gk} give $D_s(\s\to 0)=2$, but only after a transition at $D_s \approx 1.5$ for small positive $\s$. This behaviour could either be an artifact of the method used to determine the spectral dimension or something characteristic of spin foams. In any case, keeping in mind the existence of a minimal length in spin foam models, the limit $\s\to 0$ should probably be interpreted with some care.

Finally, it is easy to adapt to $d=3$ the calculation of \cite{Benedetti:2008gu} for $\k$-Minkowski and see that also in such case $D_s(\s\to 0)=2$. As the dual of $\k$-Poincar\`e algebra, this specific type of non-commutative geometry has been suggested to play a role in quantum gravity \cite{AmelinoCamelia:2003xp}. We note that, as far as short scale spectral dimension is concerned, $\k$-Minkowski agrees with CDT in $d=3$ but disagrees in $d=4$, a situation which seems to be opposite to that found about asymptotically safe gravity.

In conclusion we see that studying the spectral properties at different spacetime dimensions can lead to some surprises and provide clues on potential relations between different approaches.

%------------------------------------------------------------------------------
\section{Conclusions}
%\label{sec:3}
%------------------------------------------------------------------------------

In this paper we have presented new results about the spectral dimension in CDT. On one hand we have applied to the three-dimensional case a short-scale analysis analogous to that of AJL, and on the other we have extended the analysis to larger scales. Both parts of this work represent an important extension to the understanding of the physics of CDT models.

The short-scale results provide a hint and a first step towards an understanding of the $d$-dependence of the UV behaviour of CDT models. As explained in Sec.~\ref{sec:comparison} more and more results are being produced about the spectral dimension in quantum gravity, and a comparison among them (as functions of the dimension $d$) can be very useful.

The large-scale results represent striking evidence for classical behaviour at large scales, which goes beyond the time-dependence of the spatial volume \cite{Ambjorn:2008wc}, as the spectrum of the Laplacian contains far more detailed information than that. The least that can be said is that evidence presented here indicates that in the $N\to\infty$ limit (at fixed $\k_0$) the spectral dimension function tends to that of a stretched sphere.  By the freedom we have to rescale the ``cosmological'' proper time, we conclude that results are consistent with the production of a de Sitter ground state in the model. We note again that this spectral dimension function contains a great deal of geometrical information, which can completely characterise some simple geometries.  Ideally, we would like to say that, once the CDTs are appropriately scaled, the spectral dimension function tends to that of a sphere with standard metric.  This would provide evidence for a necessary and sufficient condition for the ground state of three-dimensional CDTs to be exactly a sphere in the classical limit.  We can establish this using the 1-to-1 relation between the spectral dimension function and spectrum of the Laplacian, and the well-known theorems \cite{Tanno1973} mentioned in section \ref{s:heat_trace} (at least up to some remaining technical concerns involving smoothness assumptions).  The comparison to the \textit{stretched} sphere prevents such a clean statement, but still gives strong evidence in favour the same conclusion.  Further, in the real world there are quantum corrections at short scales that we do not wish to remove by such a limiting procedure.  It is tempting to conjecture that, if the spectral dimension function of a geometry approximates to that of a sphere at large scales, the the geometry is ``approximately isometric to the sphere at large scales'' in some appropriate sense; this would again allow a cleaner statement.  But even without such an ambitious conjecture, the match of the spectral dimension functions exhibited in figure~\ref{f:spherematch} presents powerful new evidence for the approximation of 3D CDT simulation results to de Sitter space at large scales.

Extensions of the work could involve more extensive comparisons with the spectral dimension functions arising from other approaches to quantum gravity.
This kind of comparison could suggest ways to derive an estimation of the Planck length (or some proportional scale) from the small scale form of the spectral dimension function for CDTs, for example by relating the Planck length to the location of the maximum of $D_s(\s)$.  This would be useful if and when studies are undertaken to investigate the sub-Planckian regime of the models. 
Such studies would be necessary in order to assess the existence of a non-trivial continuum limit, $i.e.$ one with local degrees of freedom and a finite Planck length. In addition, it would be interesting to see if such limit could be taken in a way compatible with the rescaling needed to have an isotropic universe:
this would shed light on the connections between CDTs and Ho\v{r}ava-Lifshitz gravity. 

In view of the successful application of our method to 3D CDT, the same techniques can now be applied to the more physically interesting 4D CDT models.  There are no new conceptual problems to overcome to do so; the extension merely involves an application of the same procedures to the previously studied 4D simulations.  This will provide interesting new ways to explore the large scale geometry, as well as to investigate small-scale behaviour as we push simulations further into the sub-Planckian regime.

%%%%%%%%%%%%%%%%%%%%%%%%%%%%%%%%%%%%

\section*{Acknowledgements}

We are very grateful to Jan Ambj\o rn, Jerzy Jurkiewicz and Renate Loll for the use of the Monte Carlo code that was essential to this work, and also for discussions of the surrounding issues.  Thanks are also due to Pedro Machado for helpful discussions on spectral geometry which helped to lay the foundations for this work.  The Monte Carlo simulations were made possible by the facilities of the Shared Hierarchical Academic Research Computing Network (SHARCNET:www.sharcnet.ca) and Compute/Calcul Canada. Research at Perimeter Institute for Theoretical Physics is supported in part by the Government of Canada through NSERC and by the Province of Ontario through MRI.

%%%%%%%%%%%%%%%%%%%%%%%%%%%%%%%%%%%%%

\section*{Appendices}
%------------ Appendices -------------------------------------------------
\begin{appendix}
%------------------------------------------------------------------------------

%------------------------------------------------------------------------------
\section{Stretched sphere}
\label{a:rugby}
%------------------------------------------------------------------------------
\subsection{Definition and geometric properties}
%------------------------------------------------------------------------------

We define a three-dimensional stretched sphere $S^3_s$ by uniformly stretching the proper distance in the longitudinal direction. Its metric is
\be \label{metric}
ds^2_{S^3_s} = r^2 ( s^2 d\psi^2 + \sin^2\psi\ (d\theta^2 + \sin^2\theta\ d\phi^2))\, ,
\ee
where $\psi,\theta\in [ 0,\pi]$ and $\phi\in[0,2\pi]$, and $s$ is the deformation parameter (for $s=1$ we have the standard metric on a sphere of radius $r$).
It is easy to see that a manifold with such metric has conical singularities at the poles; for example, we can approximate $\sin\psi\sim\psi$ near the North pole, and by making the substitution $\{\rho = r s \psi,\varphi = \phi/s\}$ (also restricting to the subspace $\theta=\pi/2$ for simplicity),
we find the metric of a cone
\be
ds^2_0 = d\rho^2 + \rho^2 d\varphi^2\ ,
\ee
where $\varphi\in[0,2\pi/s]$.

The scalar Ricci curvature is
\be
R= \frac{ 6 \sin^2\psi +2 (s^2-1)}{r^2 s^2 \sin^2\psi}\, ,
\ee
and is divergent as expected in $\psi=0,\pi$ (incidentally its integral is finite, but that is not true for higher order invariants). On a manifold with conical singularities no closed analytical expression is known for the  coefficients of the heat kernel expansion, but we will be able to compute directly the spectrum for this particular case.

\begin{figure}[ht]
\centering \includegraphics{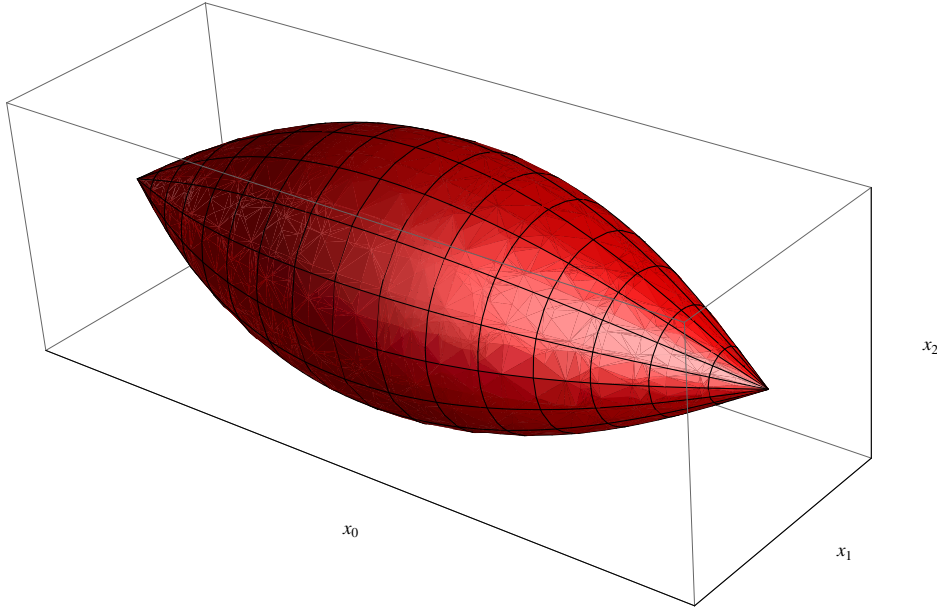}%\resizebox{6.8in}{4.5in}{  }
\caption{\small{An embedding in $\mathbb R^3$ of the $\theta=\pi/2$ section of the three-dimensional stretched sphere, with $r=1.2$ and $s=1.96$.
}\label{f:embedding}}
\end{figure}

An embedding in $\mathbb R^4$ can be obtained with the coordinate parameterization
\be
\begin{split}
 x_0 &= r \int_\psi^{\pi/2} \sqrt{s^2-\cos^2 \psi'}\ d\psi' \, ,\\
 x_1 &= r \cos\phi \sin\theta \sin\psi \, ,\\
 x_2 &= r \sin\phi \sin\theta \sin\psi \, ,\\
 x_3 &= r \cos\theta \sin\psi\, .
\end{split}
\ee
A section of such embedding is shown in Fig.\ref{f:embedding}.

%------------------------------------------------------------------------------
\subsection{The Laplacian and its spectrum}
%------------------------------------------------------------------------------

The scalar Laplacian associated to the metric \eqref{metric} is
\be
\begin{split}
\Delta_{S^3_\a} f 
            & =  \frac{1}{r^2 s^2 \sin^2\psi} \p_\psi \big(\sin^2\psi\ \p_\psi f\big)  +\frac{1}{r^2\sin^2\psi\sin\theta} \p_\theta \big(\sin\theta\ \p_\theta f\big) + \frac{1}{r^2\sin^2\psi\sin^2\theta} \p_\phi^2 f \\
            & =  \frac{1}{r^2 s^2 \sin^2\psi} \p_\psi \big(\sin^2\psi\ \p_\psi f\big)  +\frac{1}{r^2\sin^2\psi} \Delta_{S^2} f \, ,
\end{split}
\ee
where $\Delta_{S^2}$ is the Laplacian on a unit 2-sphere.
In order to look for its spectrum we start with a separation of variables, $i.e.$ we make the ansatz
\be
f(\psi,\theta,\phi) = \rho(\psi) Y_l^m(\theta,\phi)\, ,
\ee
where $Y_l^m(\theta,\phi)$ are spherical harmonics ($l=0,1,2...$ and $m=-l,-l+1,...l$), so that
\be
\Delta_{S^2} Y_l^m(\theta,\phi)\equiv  (\p_\theta^2+\frac{\cos\theta}{\sin\theta}\p_\theta+\frac{1}{\sin^2\theta}\p_\phi^2) Y_l^m(\theta,\phi) = -l(l+1) Y_l^m(\theta,\phi)\, .
\ee
As a consequence, the eigenvalue equation reduces to an ordinary differential equation for $\rho(\psi)$:
\be
\frac{1}{r^2}\left[ \frac{1}{s^2} \p_\psi^2 +\frac{2}{s^2}\frac{\cos\psi}{\sin\psi}\p_\psi -\frac{l(l+1)}{\sin^2\psi}\right] \rho(\psi) = - E\, \rho(\psi)\, .
\ee
The generic solution for this equation is found to be of the form
\be
 \rho(\psi)  = \frac{A\ P_\l^\m(\cos\psi) + B\ Q_\l^\m(\cos\psi)}{\sqrt{\sin\psi}}\, ,
\ee
where $P_\l^\m$ and $Q_\l^\m$ are the associated Legendre functions of the first and second kind, with
$\l = \frac{2\sqrt{1+s^2 r^2 E}-1}{2}$ and $\m = \frac{1}{2}\sqrt{1+4 s^2 l (l+1)}$,  and coefficients $A$ and $B$ to be fixed by boundary conditions.
Imposing regularity at $\psi=0$ we find that
\be
\frac{A}{B} = -\frac{\pi}{2} \frac{\cos \m\pi}{\sin\m\pi} \, ,
\ee
while $B$ can be fixed by normalization.

Finally, imposing regularity also at $\psi=\pi$, the eigenvalues $E$ are restricted to
\be
E_{jl} = \frac{2 j (j+1-2l +2\m ) +2l (l-1 + s^2 (l+1)-2\m) +2\m-1      }{2 s^2 r^2}\, ,
\ee
with $j\geq l$ a non-negative integer. The degeneracy of the eigenvalues is that inherited from the spectrum over $S^2$, $i.e.$
\be
D_{jl} = 2 l+1\, .
\ee

For $s=1$ the eigenvalues become independent of $l$ and the spectrum of the round 3-sphere is recovered:
\be
\begin{split}
E_{ij}^{(\a=1)} & = j (j+2)\, ,\\
D_{ij}^{(\a=1)} & = (j+1)^2\, .
\end{split}
\ee

Once we have the full spectrum the heat kernel trace can be directly evaluated by the formula
\be \label{spectral-series}
P_{S^3_s}(\s) = \frac{1}{V_{S^3_s}}\sum_{j=0}^{+\infty}\sum_{l=0}^j D_{jl} e^{-\s E_{jl}}\, ,
\ee
with the volume given by $V_{S^3_s} = 2\pi^2 r^3 s$.

%------------------------------------------------------------------------------
\section{Solution to (2+1)-dimensional $z=2$ Ho\v{r}ava-Lifshitz gravity}
\label{app:Lifshitz}
%------------------------------------------------------------------------------

A recent proposal by Petr Ho\v{r}ava \cite{Horava:2009uw} of a power-counting renormalizable theory of gravity in 3+1 dimensions has attracted lot of attention by the physics community. A possible link between this model and CDT was pointed out in \cite{Horava:2009if}, and we wish to explore it further here in light of our results.
In general, in $n+1$ dimensions, the idea is to construct an anisotropic theory of gravity, characterized by a critical exponent $z=n$, such that power-counting renormalizability is guaranteed without introducing higher-derivatives in time (which would typically spoil unitarity).
The case of our interest is hence a $z=2$ theory in 2+1 dimensions, which has already been considered in \cite{Horava:2008ih}.

We will consider here a version without detailed-balance condition and in Euclidean signature, for which the action is
\be
S = \int d t d^2 x N \sqrt{g} \left\{ \frac{2}{\k^2}\left[   \l K^2 -K_{ij}K^{ij} -2 \L + \pr{2}R  \right]  -\g \pr{2}R^2   \right\}\, ,
\ee
where $g$ is the determinant of the spatial metric\footnote{Despite being in Euclidean signature we keep using a spacetime terminology, for lack of a better one.}, $\pr{2}R$ its Ricci scalar, $N$ the lapse function, $K_{ij}$ the extrinsic curvature of the leaves of the foliation, and $K$ its trace. The coupling $\k^2$ is proportional to Newton's constant, and $\L$ is the cosmological constant, while $\l$ and $\g$ characterize the deviations from full diffeomorphism invariance ($l=1$ and $\g=0$ corresponding to general relativity in 2+1 dimensions).
In the following, since we are interested in the infrared limit of the theory, we will restrict to $\g=0$, hence leaving only to $\l\neq 1$ the characterization of the anisotropy.

The general equations of motion can be easily found and are similar to those in \cite{Lu:2009em}, but we will not report them here. We are interested in spatially spherically-symmetric solutions, and therefore we make the ansatz
\be
ds^2 = N^2(\psi)\ d\psi^2 + f(\psi) (d\th^2 + \sin^2\th\ d\phi^2)\, .
\ee
As a consequence the equations of motion reduce to
\be
\frac{2\l-1}{2N^2} \left(\frac{f'}{f}\right)^2+2\L-\frac{2}{f} = 0\, ,
\ee
\be
\frac{2\l-1}{2N}\left( \frac{1}{2} \left(\frac{f'}{f}\right)^2 + \frac{N'}{N}\frac{f'}{f}  -\frac{f''}{f} \right) -\L N = 0\, .
\ee
It is easily verified that a solution is given by
\be
f(\psi)= \frac{1}{\L} \sin^2\psi\, ,
\ee
\be
N^2(\psi) = \frac{2\l-1}{\L}\, .
\ee
This corresponds exactly to the metric \eqref{metric} for $r^2=1/\L$ and $s^2=2\l-1$.

We hence see that the large scale spectrum found in our CDT simulations in general matches the above solution to Ho\v{r}ava-Lifshitz gravity. As explained in the main text we are at this stage free to translate our results to the continuum language in such a way that the round sphere is matched, but we should keep in mind the general result when taking into account the scaling of $\k_0$ in future studies: for example, when taking the continuum limit it might turn out to be impossible to keep the Planck length finite and at the same time have a round sphere at large scales. In the renormalization group language this would mean that the coupling $\l$ does not flow to $\l=1$ in the infra-red. We hope to come back to this issue in the near future.

%------------------------------------------------------------------------------
\end{appendix}
%------------------------------------------------------------------------------

\clearpage

%------------------------------------------------------------------------------
%\begin{thebibliography}{00}
%------------------------------------------------------------------------------

%\end{thebibliography}

%\bibliographystyle{JHEP-3}
%\bibliography{CDT-refs}

\providecommand{\href}[2]{#2}\begingroup\raggedright\endgroup

\end{document}